\documentstyle[aps,preprint,epsfig]{revtex}

\def\Box{\kern1pt\vbox{\hrule height 1.2pt\hbox{\vrule width 1.2pt\hskip 3pt 
\vbox{\vskip 6pt}\hskip 3pt\vrule width 0.6pt}\hrule height 0.6pt}\kern1pt} 
\def\gtwid{\mathrel{\raise.3ex\hbox{$>$\kern-.75em\lower1ex\hbox{$\sim$}}}} 
\def\ltwid{\mathrel{\raise.3ex\hbox{$<$\kern-.75em\lower1ex\hbox{$\sim$}}}} 
\def\be{\begin{equation}} 
\def\ee{\end{equation}} 
\def\beq{\begin{eqnarray}} 
\def\eeq{\end{eqnarray}}

\begin{document} 
 
\draft
 
\title{A note on dualities in Einstein's gravity in the
presence of a non-minimally coupled scalar field}

\author{L.R. Abramo$^1$,
L. Brenig$^2$,
E. Gunzig$^{2,3}$,
Alberto Saa$^{2,4}$ \footnote{e-mails:
{
abramo@fma.if.usp.br,
lbrenig@ulb.ac.be,
egunzig@ulb.ac.be,
asaa@ime.unicamp.br}}}
 
\address{1)
Instituto de F\'\i sica, Universidade de S\~ao Paulo,
CP 66318, 05315-970 S\~ao Paulo, SP, Brazil
}
 
\address{2)
RggR, Universit\'e Libre de Bruxelles,
CP 231, 1050 Bruxelles, Belgium.
}
\address{3)
Instituts Internationaux de Chimie et de Physique Solvay,
CP 231, 1050 Bruxelles, Belgium.
}
 
\address{4)
IMECC -- UNICAMP,
C.P. 6065, 13081-970 Campinas, SP, Brazil.}

\maketitle

\begin{abstract}
We show that the action of Einstein's gravity with a 
scalar field coupled in a generic way to spacetime curvature
is invariant under a particular set of conformal transformations.
These transformations relate dual theories for which the effective 
couplings of the theory are scaled uniformly.
In the simplest case, this class of dualities reduce
to the S-duality of low-energy effective action of string theory.
\end{abstract}

 
\section{Introduction}

The theory we consider is Einstein's gravity with the addition of a 
non-minimally coupled (NMC) real scalar field:
\be
\label{L_CC} 
{\cal{L}} = \sqrt{-g} \left\{ - \frac12 F \left( \psi \right)
R +  \frac12 g^{\mu\nu} \psi_{, \mu} \psi_{, \nu} - V(\psi) \right\} \; , 
\ee 
where $F(\psi)$ is the coupling function of the scalar 
$\psi$ to spacetime curvature, $V(\psi)$ is an arbitrary 
scalar self-interaction and $8 \pi G = 1$ in our units.

It is convenient to change variables into the
``Einstein frame'', in which the action (\ref{L_CC}) reduces 
to the Einstein-Hilbert term plus a minimally coupled scalar
field \cite{Bek,ABG02}:
\be 
\label{Lt} 
L = \sqrt{-\tilde{g}} \left\{  
- \frac{1}{2} \tilde{R} +  
\frac12 \tilde{g}^{\mu\nu} \tilde\psi_{,\mu} \tilde\psi_{,\nu} 
- \tilde{V} \right\} \; .
\ee 
We will show that classical solutions $[g_1,\psi_1,V_1]$ 
of the system (\ref{L_CC}) are dual to some other
solutions $[g_2,\psi_2,V_2]$.
These dualities follow because for each Einstein-frame solution 
$[\tilde{g},\tilde{\psi},\tilde{V}]$ of system (\ref{Lt}), 
there are in general many corresponding physical solutions 
$[g_n,\psi_n,V_n]$.


\section{The Einstein frame}

Because the factor $F (\psi)$ multiplies the scalar curvature
in the Lagrangian (\ref{L_CC}), it is clear that
the physical (helicity two) degrees of freedom of the gravitational 
field appear mixed with the scalar field.
Diagonalization of the physical (or ``Jordan-frame'') action (\ref{L_CC})
can be achieved by means of a conformal transformation 
of the metric \cite{Bek}:
\be
\label{cg} 
\tilde{g}_{\mu\nu} \equiv \Omega^2 g_{\mu\nu} \; ,
\ee 
under which the Ricci scalar (in four dimensions) transforms 
as:
\be
\label{cR} 
\tilde{R} \equiv R \left[ \tilde{g} \right] =  
\Omega^{-2} \left[ R - 6 g^{\mu\nu} D_\mu D_\nu \left( \log{\Omega} \right) 
- 6 g^{\mu\nu} D_\mu \left( \log{\Omega} \right)  
D_\nu \left( \log{\Omega} \right) \right] \; , 
\ee 
where $D_\mu[g]$ are covariant derivatives.
Using this expression in Eq. (\ref{L_CC}) one obtains, after
neglecting total derivatives:
\be
\label{L2} 
\sqrt{-\tilde{g}} \left\{ -
\frac{F(\psi)}{2\Omega^2} \tilde{R} + \frac{1}{2 \Omega^2} 
\tilde{g}^{\mu\nu} \left[ \psi_{,\mu} \psi_{,\nu}
+ 6 F(\psi)
\left( {\log \Omega} \right)_{,\mu}
\left( {\log \Omega} \right)_{,\nu}
\right]
- \frac{V(\psi)}{\Omega^4} \right\} \; .
\ee

We want to rewrite this Lagrangian in such a way that it 
resembles as much as possible the Einstein-Hilbert action 
plus a minimally coupled scalar field. Hence, we take 
$\Omega^2 = \pm F(\psi)$, where the upper choice of sign
applies when $F>0$, while the minus sign
applies in the case $F<0$. This substitution leads to:
\be 
\label{L3} 
\sqrt{- \tilde{g}} \left\{ 
\mp \frac12 \tilde{R} \pm \frac12 \tilde{g}^{\mu\nu} 
\psi_{,\mu} \psi_{,\nu} \left( \frac{1}{F} + 
\frac32 \frac{F_{,\psi}^2}{F^2} \right)
- \frac{V}{F^2} \right\} \; .
\ee
Introducing a conformally transformed field and potential:
\begin{eqnarray}
\label{tpsi} 
d \tilde{\psi}^2 &\equiv& \left( \frac{1}{F} + 
\frac32 \frac{F_{,\psi}^2}{F^2} \right) \, d \psi^2 \; , 
\\
\label{tV}
\tilde{V} &\equiv& \pm \frac{V}{F^2} \; ,
\end{eqnarray}
and substituting into Eq. (\ref{L3}) we obtain: 
\be 
\label{Lf} 
{\cal{L}} = \pm \sqrt{-\tilde{g}} \left\{  - \frac12 \tilde{R} +  
\frac12 \tilde{g}^{\mu\nu} \tilde\psi_{,\mu} \tilde\psi_{,\nu} 
- \tilde{V} (\tilde\psi) \right\} \; ,
\ee 
which is, up to the global sign, simply the action for gravity with 
a minimally coupled scalar field\footnote{Since the equations of motion
are insensitive to this overall sign, the two theories, with plus
and minus signs, are identical. In any event, our main considerations 
will focus on cases where $F>0$ and the choices of sign are the
usual ones.}

On account of Eqs. (\ref{cg}), (\ref{tpsi}) and (\ref{tV}), 
the transformation is evidently singular if $F=0$.
At that point there is in fact a spacetime singularity
which splits the theory into disconnected sectors 
\cite{Star,ABGS02}.
For transformation (\ref{tpsi}) to make sense
the terms inside the square root should be positive, that is:
\be
\label{ineq}
\frac{1}{F} + \frac32 \frac{F_{,\psi}^2}{F^2} \geq 0 \; .  
\ee
If this condition is violated, then the ``effective''
gravitons and scalars have energies with opposite signs
and the theory is unstable\cite{ABG02,Sing0}.


\section{Duality}

Suppose that $\tilde\psi$ defined in Eq. (\ref{tpsi}) 
is degenerate in $\psi$, that is,
there are $n$ ($n \geq 2$) values $\psi_1, \cdots ,\psi_n$ of 
the physical field $\psi$ which correspond to the same 
value of the Einstein-frame field $\tilde\psi$.
If this is true, then there are at least two physical solutions
which are related by the conformal transformation defined by 
Eqs. (\ref{cg}), (\ref{tpsi}) and (\ref{tV}):
\be
\label{Dual}
( \, \tilde\psi \, , \, \tilde{g} \, , \, \tilde{V} )
\begin{array}{lllll} 
\quad \quad \quad ( \, \psi_1 \, , \, g_1 \, , \, V_1  \, ) \\
\quad \swarrow \\
\quad \quad \quad \quad \quad \; \Updownarrow\\
\quad \nwarrow \\
\quad \quad \quad ( \, \psi_2 \, , \, g_2 \, , \, V_2  \, ) \; .
\end{array}
\ee
The up-down arrow is a {\it duality} relation that is found by
using the conformal transformation and its inverse.
The group formed by these transformations can be discrete
(e.g., parity), continuous (e.g., scaling), or a combination
of the two.

\subsection{Case 1: $F=\zeta\psi^{2}$ }

Consider the conformal function $F(\psi)=\zeta\psi^{2}$, 
where $\zeta > 0$.
We can integrate Eq. (\ref{tpsi}) immediately, with the result:
\be
\label{tp2}
\tilde \psi (\psi) = \pm \sqrt{\frac{6 \zeta + 1}{\zeta}} 
\, \log{\left[ \frac{|\psi|}{q} \right] } \; ,
\ee
where $q$ is a positive integration constant, and the 
choice of sign is arbitrary.

\begin{figure}
\begin{center}
\epsfig{file=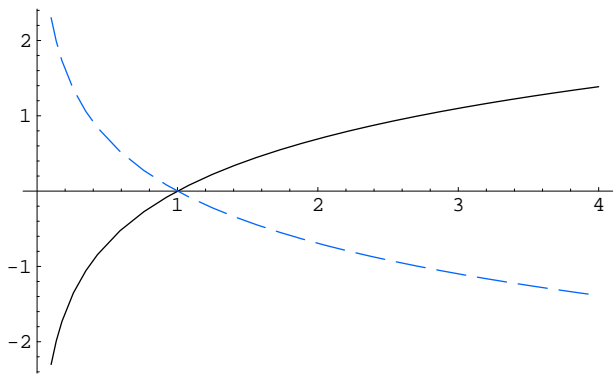,height=5.0cm}
\end{center}
\caption{Einstein-frame scalar field $\tilde\psi$ 
as a function of the physical scalar $\psi$, in the case
$F=\zeta\psi^2$.
The solid (black) and dashed (blue) lines correspond respectively to the
positive and the negative choices of the sign in Eq. (\ref{tp2}),
with $q=1$ in both cases.}
\end{figure}

The simplest type of degeneracy is the one related to the choice of 
$q$, which shifts $\tilde\psi$ up or down by a constant factor.
The dualities associated with this degeneracy are
rescalings of the scalar field by a constant, 
$\psi \rightarrow \alpha \psi$.

A more profound type of duality arises from the arbitrary choice of
the sign in Eq. (\ref{tp2}). Let's assume, without loss of generality,
that $\psi > 0$.
if $\psi_a/q_a = q_b/\psi_b$, then a differing choice of signs
in each case means that the solutions $a$ and $b$ are dual to each
other (see Fig. 1): for each $\tilde\psi$ there are two
associated physical fields, $\psi_a$ and $\psi_b$.
Given a solution of the Einstein and Klein-Gordon equations 
$\{ \psi_a , g_{\mu\nu}^{(a)} \}$ for the theory with potential $V_a$, 
there is a dual solution to those equations,
$\{ \psi_b , g_{\mu\nu}^{(b)} \}$, but with a different potential
$V_b$, which follows from Eqs. (\ref{cg}), (\ref{tpsi}) and (\ref{tV}).
Defining $q_a q_b \equiv \psi_0^2$, the dual theory is expressed in 
terms of the original one by:
\begin{eqnarray}
\label{du_p}
\psi_b &=& \frac{\psi_0^2}{\psi_a} \; , \\
\label{du_m}
g_{\mu\nu}^{(b)} &=& \frac{F(\psi_a)}{F(\psi_b)} g_{\mu\nu}^{(a)}
          = \frac{\psi_a^4}{\psi_0^2} g_{\mu\nu}^{(a)} \\
\label{du_Pot}
V_b &=& \frac{F^2(\psi_b)}{F^2(\psi_a)} V_a = 
    \frac{\psi_0^8}{\psi_a^8} V_a \; .
\end{eqnarray}

The relation between the ``effective Newton's constant'' in the dual 
theory compared with the original one is:
\begin{eqnarray}
\label{Geff_a}
G^{(a)}_{eff} &=& \frac{G}{\zeta\psi_a^2} \; ,
\\ \label{Geff_2}
G^{(b)}_{eff} &=& \frac{G}{\zeta\psi_b^2} = 
\frac{\psi_0^4}{\psi_b^4} \, G^{(a)}_{eff}
 \, ,
\end{eqnarray}
while the effective Planck masses are related by:
\be
\label{Mpeff_2}
M_{P, \; eff}^{(b)} = \frac{\psi_0^2}{\psi_a^2} M_{P, \; eff}^{(a)}
= \frac{\psi_b^2}{\psi_0^2} M_{P, \; eff}^{(a)}
\, .
\ee
If $\psi_a/\psi_0 \ll 1$, then $\psi_b/\psi_0 \gg 1$, which means that
while in the original theory theory $M_{P, \; eff}^{(a)} \gg 1$,
in the dual theory $M_{P, \; eff}^{(b)} \ll 1$ 
--- i.e., gravity is ``weak'' in the original, and ``strong'' in the
dual theory.

In fact, under duality all effective masses scale 
in the same way as the effective Planck mass.
Take the theory with scalar potential:
\be
\label{V_a}
V_a=\lambda_a M_a^4 \left( \frac{\psi_a}{M_a} \right)^{n_a} \; ,
\ee
where the coupling $\lambda_a$ is a dimensionless constant,
$M_a$ has dimensions of mass and $n_a$ is some integer.
The potential in a dual theory with $\psi_b=\psi_0^2/\psi_a$
is given by Eq. (\ref{du_Pot}):
\begin{eqnarray}
\label{V_b}
V_b &=& \frac{F_b^2}{F_a^2} V_a = 
\lambda_a \psi_0^4 \left( \frac{M_a}{\psi_0} \right)^{4-n_a} 
\left( \frac{\psi_b}{\psi_0} \right)^{8-n_a} \\ \nonumber
&=& \lambda_a \left( M_a \frac{\psi_b^2}{\psi_0^2} \right)^4
\left( \frac{\psi_b}{M_a \frac{\psi_b^2}{\psi_0^2}} \right)^{n_a} \; .
\end{eqnarray}
Therefore, we can say that duality transforms all effective 
mass scales as:
\be
\label{m_res}
M \rightarrow M_b = \frac{\psi_b^2}{\psi_0^2}  M_a \; .
\ee
However, notice that since $\psi_b$ is in fact a dynamical variable,
the dual theories are fundamentally different from each other.
From inspection of Eqs. (\ref{V_a}) and (\ref{V_b}), the exact 
statement is that duality transforms the parameters of the
potential as follows:
\begin{eqnarray} \nonumber
n &\rightarrow& n_b = 8-n_a \; ,\\ \nonumber
M &\rightarrow& \psi_0 \; , \\ \nonumber
\lambda &\rightarrow& \lambda_b = 
\lambda_a \left( \frac{M_a}{\psi_0} \right)^{4-n_a} \; .
\end{eqnarray}
The dual of the case $n=2$ with mass $M$, for instance, 
is an $n=6$ theory with coupling $\lambda M^2/\psi_0^2$
and mass scale $\psi_0$.

In the present example with $F=\zeta \psi^2$, our duality 
takes $\psi \rightarrow c/\psi$. We now show that this duality
is nothing but ``S-duality'', a symmetry of the low-energy effective 
action of string theory which switches the sign of the dilaton field,
$\phi \rightarrow -\phi$. With a conformal factor
$F(\psi)= \psi^2/4$, consider the field redefinition:
\be
\label{dil}
\psi = 2 e^{-\phi/2} \; ,
\ee
leading to the action:
\be
\label{L_st}
{\cal{L}} = \sqrt{-g} \, e^{-\phi} \, \left\{ - \frac12 R 
+ \frac12 g^{\mu\nu} \phi_{, \mu} \phi_{, \nu} - e^{\phi} V(\phi) 
\right\} \; , 
\ee 
which is just the low-energy effective action of string theory 
restricted to gravity and the dilaton. It is well known that the 
transformation $\phi \rightarrow - \phi$, together with 
an accompanying Weyl transformation of the metric, 
$g_{\mu\nu} \rightarrow \exp{(2\phi)} g_{\mu\nu}$, leaves
the action invariant --- this is the so-called S-duality 
of low-energy string theory \cite{Sdual}. By Eq. (\ref{dil}),
this duality takes $\psi \rightarrow 4/\psi$, which is
exactly the type of duality that we have discussed. Notice
that choices of $\zeta$ other than $1/4$ are obtained simply
by shifting the dilaton by a constant.

\subsection{Case 2: $F=\zeta \psi^{2k}$}

Generalizing the previous results, consider a conformal 
factor $F(\psi)= \zeta \psi^{2k}$, where $\zeta$ and $k\neq 1$
are positive real numbers. In this case the Einstein-frame field is:
\begin{eqnarray}
\label{tpsi_m}
\tilde\psi &=& \psi_0 \; \pm \; \frac{1}{k-1} \,
\sqrt{\frac{6\zeta k^2 + \psi^{2-2k}}{\zeta}} 
\\ \nonumber
& & \times 
\left[ 1 - \sqrt{6\zeta} k \left( 6\zeta k^2 + \psi^{2-2k} \right)^{-1/2}
{\rm Arcsinh} \left( \sqrt{6\zeta} \, k \, \psi^{k-1} \right) \right]  \; .
\end{eqnarray}
It is easy to show that this expression reduces to Eq. (\ref{tp2})
when $k \rightarrow 1$. The qualitative behavior of $\tilde{\psi}$ is
the same as that of Fig. 1.

In the example with $k=1$, because of the purely logarithmic dependence,
the arbitrary integration constant of $\tilde\psi$
translates into scaling invariance of the physical theory.
In the present case, the constant factor inside square 
brackets prevents the integration constant from being
absorbed into the Arcsinh. Therefore, the case $k \neq 1$ does 
not exhibit scaling invariance --- duality now involves some
transcendental relation between different values of the scalar field.
It might still be worth noting that for $k<1$ the theory regains
asymptotic scaling invariance in the limit $\psi \ll 1$.

\subsection{Case 4: $F=1-\xi\psi^2$}

This is the usual case when one considers NMC scalar fields.
For $\xi=0$ the coupling is said to be ``minimal'', while
$\xi=1/6$ corresponds to the so-called ``conformal coupling''.
This theory admits three stable sectors \cite{ABG02}: 
\begin{quote}

($a$) $\xi \leq 0$ ,

($b$) $\xi \geq 1/6$ with $\psi^2 < \xi^{-1}$ , and 

($c$) $\xi \geq 1/6$ with $\psi^2 > \xi^{-1}$ .

\end{quote}
Direct integration of Eq. (\ref{tpsi}) in this case yelds the 
following expression:
\begin{eqnarray}
\label{tp}
\tilde\psi (\psi) &=& \sqrt{\frac32} \, {\rm log} \left[
\alpha \, \left( \sqrt{\xi (6\xi-1) \psi^2} + 
\sqrt{1 + \xi (6\xi-1) \psi^2} \right)^{-\sqrt{\frac23 
\frac{6\xi-1}{\xi}}} \right. \\ \nonumber
& & \left. \times \, \frac{1+\psi \sqrt\xi}{1-\psi \sqrt\xi} \, \times 
  \, \frac{1+\sqrt{6\xi}\sqrt{1 + \xi (6\xi-1) \psi^2} 
  + \sqrt\xi (6\xi-1)\psi}{1+\sqrt{6\xi}\sqrt{1 + 
  \xi (6\xi-1) \psi^2} - \sqrt\xi (6\xi-1)\psi}
\right] \; ,
\end{eqnarray}
where $\alpha$ is an arbitrary integration constant.
We have plotted $\tilde{\psi} (\psi)$ in Fig. 2 for the
cases $\xi = -1/8$, $\xi=1/6$ and $\xi = 1$, with the
choice $\alpha=\pm 1$ to make the argument of the log 
positive (a different choice with the same sign would only 
shift $\tilde\psi$ by a constant.)

\begin{figure}
\begin{center}
\epsfig{file=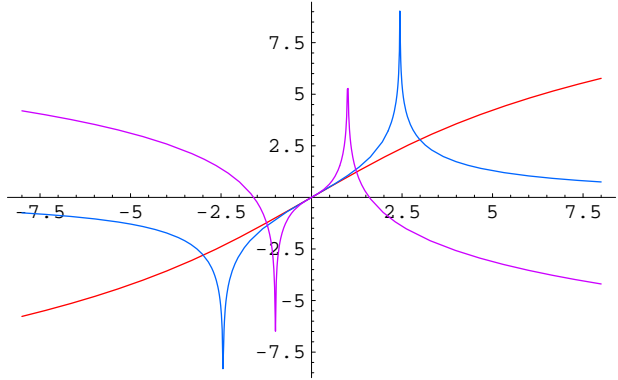,height=5.0cm}
\end{center}
\caption{Einstein-frame scalar field $\tilde\psi$ in the case
$F=1-\xi\psi^2$, as a function of the physical scalar $\psi$, 
for three values of the curvature coupling $\xi$: 
$-1/8$, $1/6$ and $1$ (red, blue and purple lines, respectively.)}
\end{figure}

The case $\xi = -1/8$, representing sector $a$ ($\xi \leq 0$), 
is the simplest: $\tilde\psi$ is monotonic in $\psi$. In this
case there is no degeneracy and, hence, no duality.

The case $\xi=1/6$ (the middle line at the sides of Fig. 2) 
is a threshold case: here, the Einstein-frame
field $\tilde\psi$ goes to zero when $|\psi| \rightarrow \infty$,
so for each value of $\tilde\psi$ there are two associated
values of the physical field $\psi$ --- one in sector $b$,
$|\psi_b| < \sqrt{6}$ and $F(\psi_b) > 0$, and one in sector 
$c$, $|\psi_c| > \sqrt{6}$ and $F(\psi_c) < 0$.
In this case expression (\ref{tp}) simplifies considerably:
\be
\label{tp_c}
\tilde{\psi} (\psi) \, = \, \left\{ 
\begin{array}{lll} 
\sqrt{6} \tanh^{-1}{[\psi_b/\sqrt{6}]} \quad , 
\quad  \psi_b^2 < 6  \quad (\alpha=+1) \; , \\
\\
\sqrt{6} \tanh^{-1}{[\sqrt{6}/\psi_c]} \quad , 
\quad  \psi_c^2 > 6  \quad (\alpha=-1) \; .
\end{array} \right. 
\ee
For $\xi=1/6$ the explicit relation between the two physical
solutions $(g_b,\psi_b,V_b)$ and $(g_c,\psi_c,V_c)$
corresponding to the same Einstein-frame solution
$(\tilde{g},\tilde\psi,\tilde{V})$
can be obtained by solving for the arguments of the $\tanh^{-1}$,
and by using Eqs. (\ref{cg}) and (\ref{tV}):
\begin{eqnarray}
\label{du_c_p}
\psi_b \, \psi_c &=& 6 \; , \\
\label{du_c_g}
\psi_b \, g_{\mu\nu}^b &=& \psi_c \, g_{\mu\nu}^c \; , \\
\label{du_c_V}
\psi_b^2 \, V_b &=& - \, \psi_c^2 \, V_c \; .
\end{eqnarray}
It can be easily seen from Eq. (\ref{du_c_p}) that this particular
duality relates physical solutions $\psi_b$ and $\psi_c$ such
that, for $F_b \rightarrow 0^+$, $F_c \rightarrow 0^-$, 
and for $F_b \rightarrow 1$, $F_c \rightarrow -\infty$.

The case $\xi > 1/6$ is more interesting.
In the example of Fig. 2, of $\xi = 1$, the  region $|\psi| < 1$ 
(between the poles) corresponds to our sector $b$, 
while region $|\psi| > 1$ corresponds to our sector $c$. 
As before, the choice of the integration constant was
$\alpha = \pm 1$ to make the argument of the $\log$ in Eq. (\ref{tp}) 
positive.

As can be easily seen from Fig. 2, when $\xi>1/6$,
for each value of $\tilde\psi$ there are now {\it three} associated 
values of $\psi$ --- one in sector $b$ and one in each of the two
independent branches of sector $c$.

Explicit relations between the dual solutions are found by solving
for the threefold degeneracy of Eq. (\ref{tp}).
Unfortunately, it has not been possible to
solve these equations exactly, except in the case $\xi=1/6$
explained above. Nevertheless, it suffices for our purposes to 
inspect the asymptotic limits
$|\psi_1| \sqrt\xi \rightarrow 1$ and $|\psi_2| \rightarrow \infty$.
We find that for each $\tilde\psi$ there is a 
threefold degeneracy:
\be
\label{as_du}
1 - |\psi_{b1}| \sqrt\xi \, \approx \, 
|\psi_{c1}| \sqrt\xi - 1 \, \approx \, 
\beta \, \left[ -{\rm sign}\left( \psi_{c1} \right) \psi_{c2} 
\right]^{-\sqrt{\frac32 \frac{6\xi-1}{\xi}}} \,
\ee
where $\beta$ is a $\xi$-dependent constant,
$|\psi_{b1}| \sqrt\xi \rightarrow 1^-$,
$|\psi_{c1}| \sqrt\xi \rightarrow 1^+$ and
$|\psi_{c2}| \rightarrow \infty$.
One of the degenerate solutions belongs to sector $b$, while 
the other two belong to sector $c$, as is also clear from Fig. 2.
However, two of the dual solutions now have $|F(\psi_1)| \rightarrow 0$
in the limit $|\psi_1| \sqrt\xi \rightarrow 1$, while the third 
solution $|\psi_c| \rightarrow \infty$, which means that
$F(\psi_c) \rightarrow -\infty$.

\section{Conclusion} 

We have found a class of dualities of Einstein's gravity with a 
non-minimally coupled scalar field.
The dualities relate inequivalent theories for which the effective 
couplings scale by the conformal factor $F(\psi)$. In particular,
the dualities relate theories for which gravity is weak with theories 
for which it is strong. In the simplest case, $F(\psi)=\zeta\psi^2$, 
these dualities reduce to the S-duality of low-energy effective 
string theory. We have also studied the cases $F=\zeta \psi^{2k}$ and
$F=1-\xi \psi^2$, and have shown how dualities are manifested 
in those theories.

\acknowledgments 

We would like to thank E. Abdalla, R. Brandenberger, 
N. Berkovitz and V. Rivelles 
for valuable discussions. L. R. A. and A. S. thank the 
Physics Department of the Universit\'e Libre de Bruxelles 
for its hospitality during parts of this project.
The authors would like to acknowledge support from 
the EEC Contract \# HPHA-CT-2000-00015, from
Association ``Science et Environnemen'' (Mallemort, France)
and from the OLAM Fondation Pour La Recherche Fondamentale 
(Brussels).

\end{document}